# CLIC Background Studies and optimization of the innermost tracker elements

D. Dannheim[1], A. Sailer[1], J. Trenado[2*], M. Vos[3]

1 – European Organization for Nuclear Research (CERN)
CH-1211 Genève 23 – Switzerland

2 – University of Barcelona (UB) – Dep. of Structure and Matter Constituents
Martí i Franquès 1, 08028 Barcelona – Spain

2 – Instituto de Física Corpuscular (IFIC)
Catedrático José Beltrán 2, 46980 Paterna – Spain

* corresponding author: jtrenado@ecm.ub.es

The harsh machine background at the Compact Linear Collider (CLIC) forms a strong constraint on the design of the innermost part of the tracker. For the CLIC Conceptual Design Report, the detector concepts developed for the International Linear Collider (ILC) were adapted to the CLIC environment. We present the new layout for the Vertex Detector and the Forward Tracking Disks of the CLIC detector concepts, as well as the background levels in these detectors. We also study the dependence of the background rates on technology parameters like thickness of the active layer and detection threshold.

## 1  Introduction

The Compact Linear Collider (CLIC) [1] is one of the proposals for an electron-positron collider to explore the energy frontier. The conceptual design develops an $e^+ e^-$ collider with two main linacs of 21.02km, where electrons and positrons will be accelerated at 100MV/m to produce collisions with a nominal center-of-mass energy of 3TeV.

The detector concepts ILD (International Large Detector) [2] and SiD (Silicon Detector) [3], originally designed for the ILC, form the starting point for the design of the CLIC detectors. For the CLIC CDR [4] the ILD and SiD design were adapted to cope with the much larger center-of-mass energy and the extremely short bunch spacing. Early studies of machine backgrounds in the ILD concept [5] show that at a center-of-mass energy of 3 TeV the hit density due to beam-induced background is significantly higher than at the ILC. We reassess the background levels at CLIC. The tracking systems that are most strongly affected are those situated at small distance to the interaction point, and the innermost radii of the forward tracking system. We therefore redesign the barrel VXD (Vertex Detector) and the FTD (Forward Tracking Disks) to keep manageable background levels in all systems.

The background studies assume generic solid-state detectors. Candidate detector technologies for the vertex detector and tracking disks show strong variations in a number of parameters that can have a strong impact on the background hit density. We consider two parameters. The thickness of the active layer has a strong impact on the background hit density, as particles frequently impinge on the detectors under a very shallow angle. The active thickness varies from as little as 500 nm in Geiger-mode devices [6] to 200-300 µm in mainstream hybrid pixel detector technology employed in the LHC experiments. The second parameter, the detection threshold, is typically related to the former. Geiger-mode devices are sensitive to single electrons, while hybrid pixel technologies are insensitive to ionization signal below several hundreds of charge carriers. Therefore these characteristics could lead to important variations in the background hit density.

In Section 2 of this contribution we present the layout of the VXD and FTD in the CLIC detector concepts and discuss the resulting background levels. The impact on these results of typical pixel detector technology choices is studied in Section 3.

## 2 FTD & VXD new layout

Early studies show that the contribution of incoherent pair production to the hit density in the tracking elements at low radius at CLIC is significantly increased with respect to the ILC [4]. The production of hadrons due to photon-photon fusion ($\gamma\gamma \rightarrow$ hadrons) is moreover found to contribute much more significantly at CLIC [7]. The charged particles from this process tend to be emitted in the forward region. Pair background produces very soft particles that do not reach large radii in the strong magnetic field of the tracking volume. The momentum spectrum of charged particles due to $\gamma\gamma \rightarrow$ hadrons is much harder and this process is the dominant source of background at large radius.

The primary handle on the background hit density due to the dominant pair production process is the inner radius of the vertex detector and the innermost tracking disks. To obtain the same levels of background as in the ILC experiments, it has to be increased significantly.

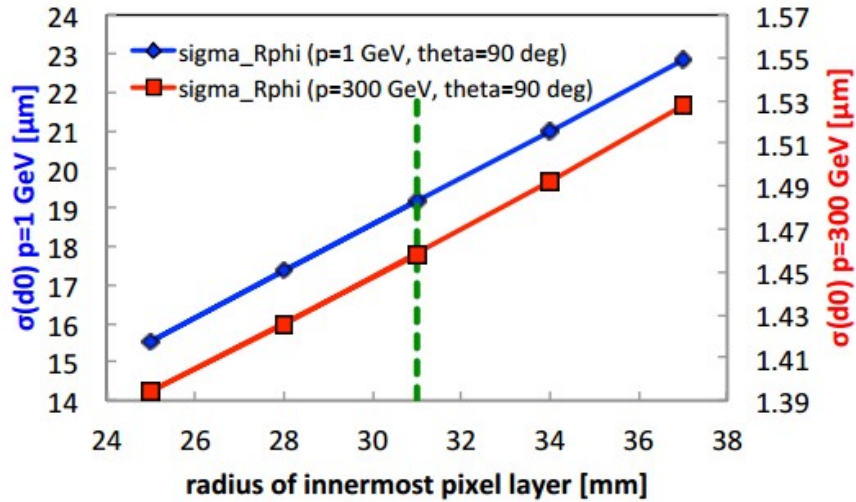

Figure 1: The dependence of the transverse impact parameter resolution on the inner radius of the vertex detector barrel, that is varied from 25 to 37 mm (the ILC detector concepts have inner radii of ~15 mm).

A change in this parameter has an impact on the precision of the extrapolation of charged particle tracks to the interaction point. The impact parameter resolution versus inner radius of the barrel vertex detector is shown in Figure 1.

The impact parameter resolution is usually expressed as a squared sum of a constant term a that depends on the single-point resolution of the detector and a "material" or "multiple scattering" term b that depends on the momentum and polar angle of the charged particle.

$$\sigma_{R\Phi} = \sqrt{a^2 + \frac{b^2 GeV^2 \sin^3\theta}{p_T^2}} \qquad (1)$$

The constant term a is quite insensitive to the inner radius; the impact parameter resolution for high momentum tracks varies from 1.4 to 1.5 μm, well below the goal of a≈5 μm, for the radii considered here. The "material" term does show a significant dependence on the inner radius, as shown by the variation in the resolution for particles with p = 1 GeV, that is dominated by this term.

FTD and Beam Pipe was redesign to reduce background levels. The conical section of the beam pipe was made thick, to reduce backscatters. The conical part of the beam pipe was moved outside the tracking acceptance, making it pointing to the IP and one disk layer was added to reduce the lever arm for the track extrapolation.
In the new layout of the VXD and FTD shown in Figure 2, the innermost layer of the barrel vertex detector is placed at 31mm from the impact point, to yield a background hit density per readout cycle that is comparable to the ILC [8]. Of several layout variants a detector design with 3 double layers was chosen. As a consequence, the specification on the "material" term b of the impact parameter resolution is relaxed to 15 μm.
This modification of the barrel vertex detector design has an impact on FTD as well. To keep good polar angle coverage and provide a first measurement point as close as possible to the interaction point, three double-layer disks are located close to the end of the barrel vertex detector. The CLIC detector thus has 11 disks covering from 6.6 to 32.5 degrees, compared to 7 disks covering 5 to 36 degrees in the [5] ILD concept at the ILC. This more powerful forward tracking system helps cope with the increased importance of the forward region in a multi-TeV collider [9].

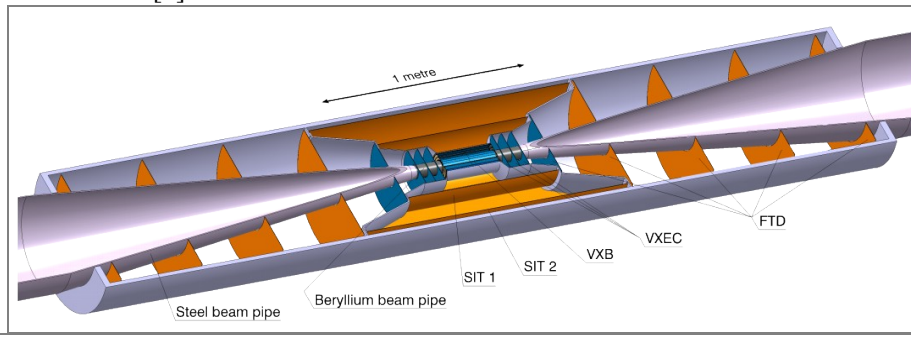

Figure 2: Sketch of tracking detectors in CLIC_ILD.

With the adapted geometry we proceed to determine the background hit density. A map on the r-z plane of the background levels of the detector elements in the innermost part of the tracking volume is shown in Figure 3. Results were obtained with 50μm thickness for the pixel sensors with a threshold of 3.4keV and 275μm thickness for the strip sensors with threshold of 17keV

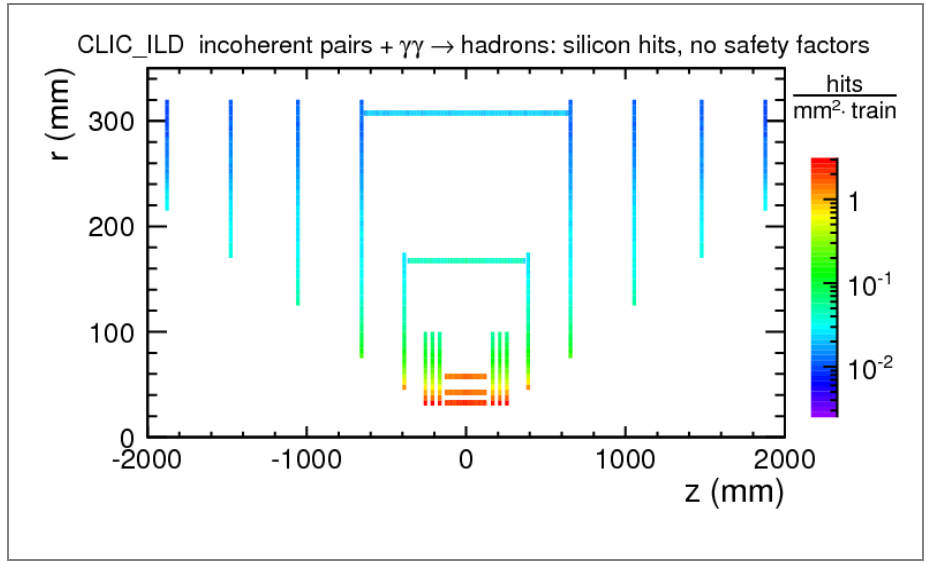

Figure 3: Average hit density in the tracking detectors from incoherent electron positron pairs and γγ→hadrons.

The contributions from incoherent pair production and γγ→hadrons are added. The former clearly dominates the hit density in the innermost layers, while the contributions are comparable for the largest radii considered here.
No safety factor is applied to account for the uncertainty in the rate. In the CLIC CDR a safety factor of five is considered for the contribution from pair production and of two for the one from γγ→hadrons. Hits are counted as GEANT4 energy deposits and the number of pixels with a signal above threshold in the cluster created by each particle is not accounted for. The next Section addresses the dependence of the result on a specific technology choice, that must be taken into account to reach a more realistic estimate.

## 3  Technology Dependence

Most background studies consider a generic solid state vertex detector technology, with a sensitive detector thickness (50 μm) and threshold (any particle producing an energy deposit is counted) that are considered to be typical of the candidate detector technologies. In the present technologies a rather broad range is found for energy threshold can be from single-electron detection to thresholds of 400e, while the thickness of the active volume lies between 1μm and 200μm. These differences lead to different hit density in the detector. A reduction of the sensitive thickness leads to much smaller clusters for inclined tracks. The technologies with the thinnest sensitive layer must have a low detection threshold because of the reduction of the energy deposition by Minimum Ionizing Particles (for Silicon, on average, approximately 80 electron-hole pairs are created for each micron the charged particle travels through the sensitive material). In this section we investigate how the hit density depends on these parameters.

Detectors with single-electron sensitivity are expected to detect also very small energy deposits from photons, for instance due to Compton scattering. Thus, such technologies could be very sensitive to the CLIC photon background. We therefore study the hit density due to photons in more detail.

These studies were done using GEANT4 [10] based full detector simulation in MOKKA [11] and Marlin [12], with CLIC_ILD model using QGSP_BERT_HP and QGSP_BIC physics list. The pair background corresponding to one bunch crossing was used with range-cut equal to 5μm. The default cut in Mokka on the energy of particles (the TPC-cut) is set to 0 eV. Hits due to photons were required to release at least 3.6eV, the minimum energy necessary to generate an electron-hole pair in silicon [13].

Fig. 4 shows the photon spectrum in the FTD. This includes photons traversing the detector without leaving any energy deposit. An important spectrum above 10keV is populated by photons that come directly from the beam and hit the detector within 1 ns after the bunch crossing. The soft spectrum, below 10eV, corresponds to photons reflected from the very forward calorimeter. This last source is not important, as the energy deposits are too small to be detected in Silicon.

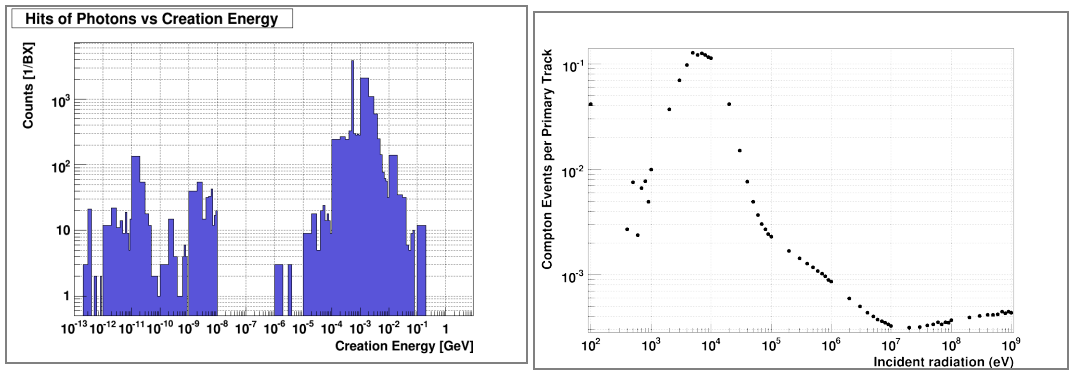

Figure 4. Leftmost panel: Spectrum of hits of photons with any deposited energy in the tracking layers, including null deposition. Rightmost panel: Number of Compton events expected per primary track in 50μm of silicon plus 8μm of $SiO_2$.

Our simulation shows that only approximately one per mil of the hits in the FTD is due to photons, even for the lowest detection threshold considered here. This can be understood considering the Compton cross-section in Fig. 4. The probability that a photon deposits energy in a thin Silicon layer is quite substantial for photons in the keV range. However, For photon energies of approximately 1 MeV, where an important population exists, the Compton cross-section is negligible.

To study the effect on the hit density and cluster size of detector thickness and detection threshold, the disks of the FTD were divided in 25μm x 25μm areas. Detailed information was retrieved from GEANT4, so multiple energy deposits in small steps, along the tracks inside the disks, were projected on the pixel grid, and pixels with energy depositions above threshold were counted as "hit" pixels.

For technologies with very thin layers the cluster size for inclined tracks is significantly reduced. On the other hand, technologies with low threshold are expected to see an increase in the number of hits due to low-energy deposits.

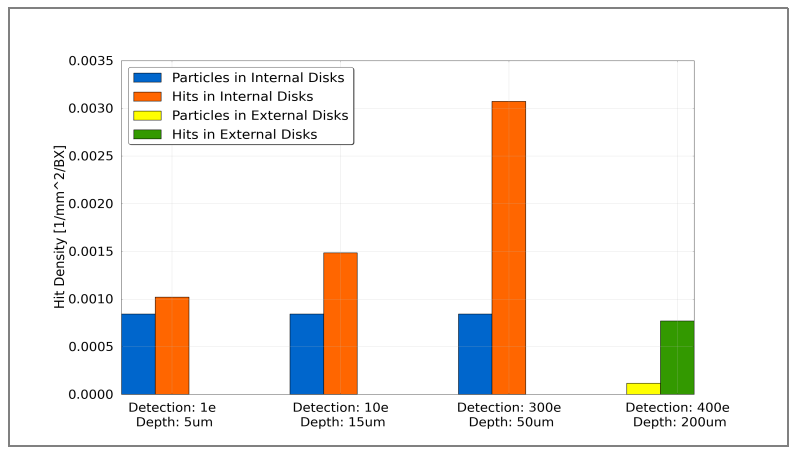

Figure 5: Comparative of hit density for different technologies.

The net result of the two competing effects is shown in Figure 5, for a number of representative combinations of detector thickness and detection threshold. The number of particles leaving a signal in the innermost FTD (the blue bars) remains roughly constant for the three combinations considered here. The smaller detection threshold leads to a negligible increase in the number of hits.
The number of "hit" pixels, however, increases strongly with increasing thickness of the sensitive layer. Charge sharing due to inclined tracks in devices with relatively thick active layers (in relation to the small pixel size) can thus lead to a significant increase in the occupancy.

## 4  Conclusions

We present the new layout for the Vertex Detector and Forward Tracking Disks in the CLIC detector concepts. The most significant adaptation to the CLIC environment is the modification of the inner radius of the vertex detector, that is located at 31mm from the interaction point. As a consequence, the specification for the multiple scattering term b of the transverse impact parameter is relaxed to 15μm. To guarantee robust coverage in the forward region the FTD is equipped with 11 disks, six of which are on double-layer disks as close as possible to the barrel vertex detector. We also present the hit density due to machine background on the innermost tracker elements.

Different vertex detector technologies considered for CLIC span a broad range of detector thicknesses and detection thresholds. Simulations taking into account both parameters show that the detection threshold (down to single-electron sensitivity) is of relatively minor importance. In particular, the contribution of photons to the hit density is at the per mil level. The detector thickness, on the other hand, strongly affects the occupancy. As background particles frequently hit the detector under a very shallow angle, the average cluster size grows approximately linearly with increasing detector thickness, leading to important differences between the thinnest and thickest detector technologies on the market.

## 5  References

[1] J. Ellis, I. Wilson, Nature, 409, (2001).

[2] Toshinori Abe, et al., ILD Concept Group - Linear Collider Collaboration. FERMILAB-LOI-2010-03, FERMILAB-PUB-09-682-E, DESY-2009-87, KEK-REPORT-2009-6, Feb 2010. 189pp
[3] H. Aihara, et al., SiD SLAC-R-944, May 26, 2010. 156pp
[4] CLIC Conceptual Design Report, ANL-HEP-TR-12-01, CERN-2012-003, DESY-12-008, KEK report 2011-7, https://edms.cern.ch/document/1180032/
[5] J.Brau, et al. ILC Reference Design Report Volume 1 – Executive Summary. 2007.
[6] E. Vilella et al., these proceedings
[7] A. Sailer, these proceedings, D. Dannheim and A. Sailer. Beam-induced Backgrounds in the CLIC Detectors. LCD-Note-2011-021
[8] D. Dannheim and M. Vos. Layout simulation studies for the vertex and tracking region of the CLIC detectors, LCD-Note-2011-031
[9] J. Fuster et al., Forward tracking at the next e+e- collider, Part I: the physics case, JINST 4 P08002 (2009)
[10]Nuclear Instruments and Methods in Physics Research, A 506 (2003) 250-303.
[11]P. Mora de Freitas and H. Videau, "Detector simulation with MOKKA / GEANT4: Present and future". Prepared for International Workshop on Linear Colliders (LCWS 2002), Jeju Island, Korea, 26-30 Aug 2002.

[12]Nuclear Instruments and Methods in Physics Research, A 559 (2006) 177-180
[13]S.M. Sze, Kwok K. NG, Physics of Semiconductor Devices. (2007)